# A Case For Amplify-Forward Relaying in the Block-Fading Multi-Access Channel

Deqiang Chen, Kambiz Azarian,
J. Nicholas Laneman, *Member, IEEE*

**Abstract**

This paper demonstrates the significant gains that multi-access users can achieve from *sharing* a single amplify-forward relay in slow fading environments. The proposed protocol, namely the multi-access relay amplify-forward, allows for a low-complexity relay and achieves the optimal diversity-multiplexing trade-off at high multiplexing gains. Analysis of the protocol reveals that it uniformly dominates the compress-forward strategy and further outperforms the dynamic decode-forward protocol at high multiplexing gains. An interesting feature of the proposed protocol is that, at high multiplexing gains, it resembles a multiple-input single-output system, and at low multiplexing gains, it provides each user with the same diversity-multiplexing trade-off as if there is no contention for the relay from the other users.

## I. Introduction

### A. Motivation

In recent years, cooperative communications has received significant interest (e.g., [1]–[7]) as a means of providing spatial diversity for applications in which temporal, spectral, and antenna diversity are limited by delay, bandwidth, and terminal size constraints, respectively. Cooperative techniques offer diversity by enabling users to utilize one another's resources such as antennas, power, and bandwidth. As a consequence, most cooperative protocols share the characteristic

This work has been supported in part by NSF through Grant CCF05-15012.

Deqiang Chen, Kambiz Azarian and J. Nicholas Laneman are with the Department of Electrical Engineering, University of Notre Dame, Notre Dame, IN 46556, Email: {dchen2, kazarian, jnl}@nd.edu

that they require substantial coordination among the users. In a wireless setting, establishing this level of user cooperation may be impractical due to cost and complexity considerations. Inspired by this observation, the current paper focuses on an alternative architecture, namely, the multi-access relay channel (MARC) [4], [8] and proposes a strategy called the multi-access amplify-forward (MAF) that allows the users to operate as if in a normal (non-cooperative) multi-access channel. In this system, the users need not be aware of the existence of the relay, *i.e.*, all cost and complexity of exploiting cooperative diversity is placed in the relay and destination. Such an architecture may be suitable for infrastructure networks, in which the relay and destination correspond respectively to a relay station and a base station deployed and managed by the service provider. It is worth noting that since a single relay is *shared* by multiple users in the MARC, the extra cost of adding the relay is amortized across many users and may thus be more affordable, especially as the number of users in the system grows. Thus, this approach facilitates a graceful transition from existing systems to cooperative ones.

### *B. Related Research*

In this section, we provide a brief review of the related research. The MARC was first introduced in [8] as a model for topologies in which multiple sources communicate with a single destination in the presence of a relay. Information-theoretical treatment of the MARC has focused on two aspects, namely, the capacity region and the diversity-multiplexing tradeoff (DMT). Using a partial-decode-forward strategy, [7] compares the AWGN MARC with cooperative multi-access communications and shows that the former achieves higher rates than the latter. Using a full-duplex relay, [4] shows that a decode-forward strategy achieves the capacity of AWGN MARC assuming the relay is geometrically close to the sources. For the general MARC, however, the optimum relaying strategy (in terms of achieving the ergodic capacity) remains unknown.

The DMT of the MARC [1] is studied in [5], [6]. In [5], the DDF strategy is applied to the MARC. In DDF, the relay does not decode until it collects sufficient information for error-free detection of the message. It then re-encodes the message and sends it over the remaining portion of the time. For the MARC, DDF is shown to achieve the optimal DMT for low multiplexing gains. However, at high multiplexing gains, it becomes suboptimal. Another relaying strategy for the MARC is compress-forward (CF) [6]. In CF, the relay employs Wyner-Ziv coding to compress its received signal and forward it to the destination. The CF achieves the optimal DMT at high multiplexing gains [6], but suffers from significant diversity loss for low multiplexing gains.

[1] In the rest of paper, we focus on the block fading scenario and the term “MARC” refers to the “block-fading MARC”.

### *C. Summary of Results*

This section summarizes our contributions. Assuming a half-duplex relay, we propose a MAF protocol for MARC and demonstrate significant gains that it brings to multi-access users. Since MAF is essentially an amplify-forward (AF) protocol, the relay does not require complicated decoding and encoding. In contrast, some of the previously proposed MARC protocols, such as dynamic decode-forward (DDF) [5] or compress-forward (CF) [6], require complex signal processing at the relay. The benefits of the proposed protocol do not limit to complexity aspects. As argued in the sequel, the MAF protocol not only uniformly dominates the CF protocol, but also outperforms the DDF protocol in the high multiplexing regime. More specifically, MAF achieves the optimal diversity-multiplexing trade-off (DMT) [9] of the MARC for multiplexing gains greater than $1/3$. This is somewhat counterintuitive considering the fact that the AF relay protocols generally suffer from a significant performance loss in the high multiplexing regime [2], [3]. It is also worth noting that each user in the MAF protocol takes the same benefit from the relay as if it was the only user present, *i.e.*, the advantage of using a single relay does *not* vanish as the number of users grows. Overall, MAF provides a nice balance between complexity and performance.

## II. Model and Protocol

### *A. Notation*

In this paper, random variables are denoted using the sans serif font (*e.g.*, $\mathsf{x}$) while random vectors are denoted with bold sans serif (*e.g.*, $\mathbf{\mathsf{X}}$). Calligraphic letters denote events or sets (*e.g.*, $\mathcal{S}$).

### *B. Model*

The MARC is distinguished from the standard multi-access channel by the existence of one or more relays solely intended to facilitate communication between the users and the destination.

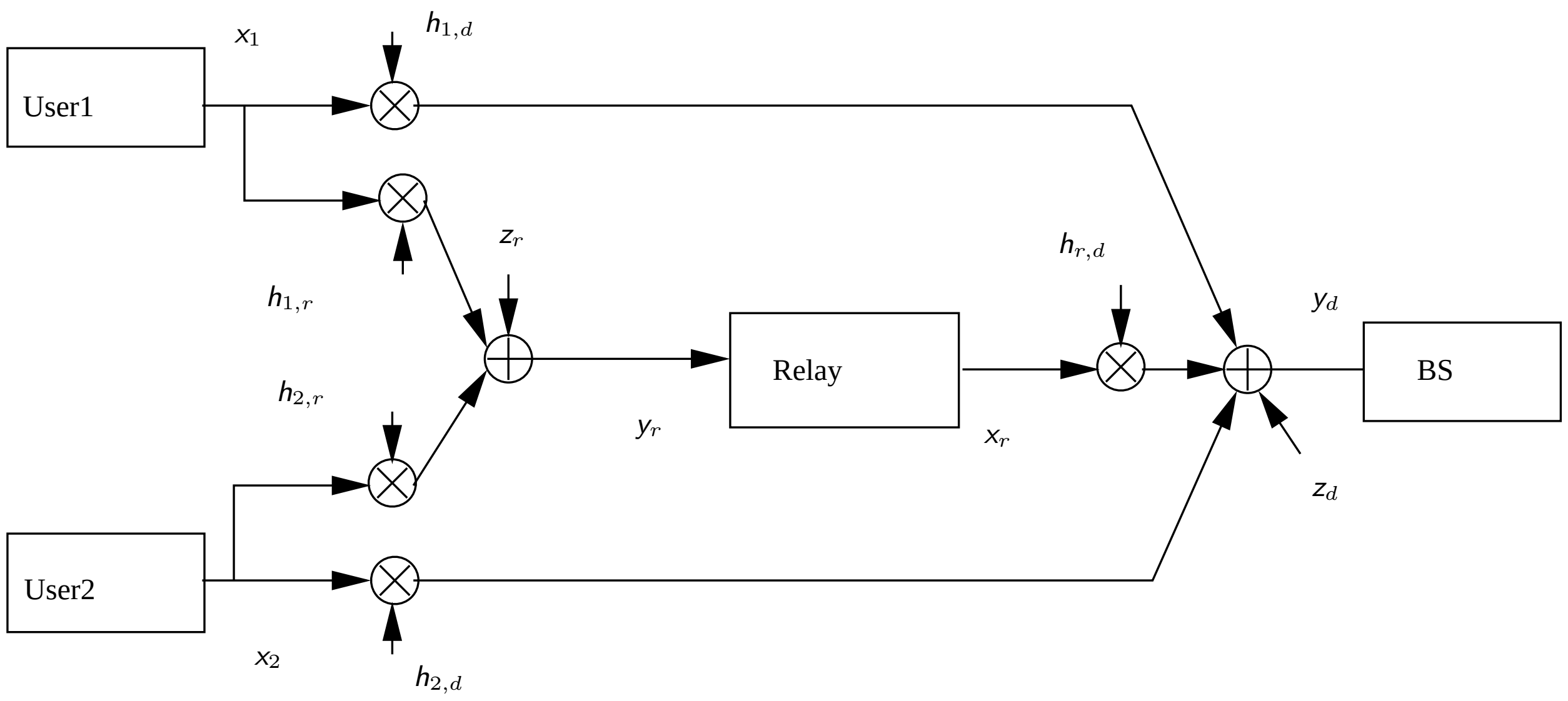

Fig. 1. Multi-access relay channel (MARC) with two users, multiplicative fading, and additive noise.

For simplicity of presentation, this paper focuses on the case of two users and one relay as shown in Fig. 1.

All wireless links are assumed to be frequency non-selective, Rayleigh block fading channels. As terminals are in different locations, fading coefficients of different links are assumed to be independent. Moreover, the channel fading coefficients remain constant within a block of $l$ symbols, but change independently from one block to the other. The block length $l$ is assumed to be long enough such that channel state information (CSI) can be tracked at the receiving end of each link, but not be available to or otherwise not exploited by the transmitting end. Furthermore, we consider the scenario in which the destination has knowledge of all CSI, including those of the user-relay links. Without loss of generality in the analysis of DMT, we assume channel fading coefficients are complex Gaussian random variables with zero mean and unit variance and the variance of the AWGN is also taken to be unity.

In order to characterize the performance of the proposed protocol in the high SNR regime, the DMT is adopted as the performance metric [9]. This paper mainly focuses on the symmetric case, *i.e.*, the two users transmit their messages at the same data rate of $R/2$ bits per channel use (bpcu). Furthermore, the two users and the relay use the same transmission power $\rho$. We consider a family of codes $\mathcal{C}(\rho) = \{\mathcal{C}_2(\rho), \mathcal{C}_2(\rho)\}$ indexed by SNR $\rho$, such that User $i$'s codebook $\mathcal{C}_i(\rho)$ has a data rate $R(\rho)/2$ and block length $l$. We consider a joint ML decoder at the base station and denote the error probability as $P_{\mathcal{E}}(\rho)$. For this family of codes and decoding schemes, we define the multiplexing gain and diversity gain as

$$r_t := \lim_{\rho\to\infty} \frac{R(\rho)}{\log\rho} \quad , \quad d := \lim_{\rho\to\infty} -\frac{\log P_{\mathcal{E}}(\rho)}{\log\rho}.$$

The individual multiplexing gains for User 1 and 2 are similarly defined and denoted by $r_1$ and $r_2$, respectively, with $r_1 = r_2 = r_t/2$.

### *C. Multi-Access Amplify-Forward*

Next, we describe the proposed MAF protocol. In MAF, the relay listens to the two users during the first half of the block; then, in the second half of the block, it simply amplifies and broadcasts the signal it received in the first half. The two users both continue transmitting their messages throughout the block. During the first half of the block, the equivalent channels seen by the destination and the relay are

$$y_d[j] = \sum_{i=1}^{2} h_{i,d} x_i[j] + z_d[j], \tag{1}$$

$$y_r[j] = \sum_{i=1}^{2} h_{i,r} x_i[j] + z_r[j], \tag{2}$$

respectively, where: $j \le l/2$ denotes the time index; $h_{i,d}$ and $h_{i,r}$ denote the fading coefficients of the user $i$-destination and user $i$-relay links, respectively; and $x_i$ denotes the signal transmitted by user $i$. Likewise, the equivalent channel seen by the destination during the second half of the block is

$$y_d[j] = \sum_{i=1}^{2} h_{i,d} x_i[j] + h_{r,d} x_r[j] + z_d[j] \tag{3}$$

for $l/2 \le j \le l$, where $h_{r,d}$ denotes the fading coefficient of the relay-destination link, and $x_r$ denotes the signal transmitted by the relay. Note that

$$x_r[j] = b y_r[j - l/2] \quad \text{for} \quad l/2 < j \le l,$$

where $b$ denotes the relay's amplification coefficient, which is chosen, *e.g.*, to minimize the outage probability at the target data rate and SNR, subject to the relay's transmission power constraint, *i.e.*,

$$|b|^2 \le \frac{\rho}{\sum_{i=1}^{2} |h_{i,r}|^2 \rho + 1}.$$

The base station is assumed to know the amplification coefficient $b$ for decoding the two messages. Note that in the single user scenario, the MAF protocol reduces to the non-orthogonal amplify-forward (NAF) protocol of [3].

The channel expressed in (1) and (3) can be regarded as a multiaccess channel with multiple transmit and receive antennas. However, the channel matrix is asymmetric and the inputs are correlated.

Since the users may not be aware of the existence of the relay, we assume that each user simply uses the capacity-achieving codebook for the corresponding MAC, *i.e.*, each codebook consists of i.i.d complex Gaussian random variables. Such inputs need not be optimal for MAF in terms of capacity or outage probability, due to the correlation that exists between the relay's signal and those of the users. However, in terms of DMT, Gaussian input turns out to be optimal for MAF at high multiplexing gains.

The following theorem provides the DMT for MAF.

*Theorem 1:* For the symmetric MARC with two users and one relay, the DMT of the MAF protocol for $0 \leq r_1 \leq 1/2$ is given by

$$d_{MAF}(r_1) = \begin{cases} 2 - 3r_1, & \text{for } 0 \leq r_1 \leq \frac{1}{3} \\ 3(1 - 2r_1), & \text{for } \frac{1}{3} \leq r_1 \leq \frac{1}{2} \end{cases}, \tag{4}$$

where $r_1$ denotes each user's multiplexing gain.

*Proof:* On one hand, the proof uses the machinery of Theorem $2$ in [10] and Lemma $2$ in [5], and on the other hand adopts some of the techniques of Theorem $3$ in [3]. Therefore, we only provide a sketch of the main steps involved and focus on the novel parts.

Following the outline of [10] and [5], we upper bound the joint error probability at the destination, with the sum of the so-called type-$\mathcal{S}$ error probabilities, *i.e.*, the probability that the destination makes errors in decoding the users in set $\mathcal{S}$, assuming the rest were decoded correctly. For the two user MARC, we have

$$P_{\mathcal{E}} \leq P_{\mathcal{E}_{\mathcal{A}}} + P_{\mathcal{E}_{\mathcal{B}}} + P_{\mathcal{E}_{\mathcal{C}}}, \tag{5}$$

where $\mathcal{E}_{\mathcal{A}}$, $\mathcal{E}_{\mathcal{B}}$ and $\mathcal{E}_{\mathcal{C}}$ are the type-$\mathcal{S}$ error events corresponding to $\mathcal{S}$ being $\{1\}$, $\{2\}$ and $\{1, 2\}$, respectively. To characterize $P_{\mathcal{E}_{\mathcal{A}}}$, $P_{\mathcal{E}_{\mathcal{B}}}$ and $P_{\mathcal{E}_{\mathcal{C}}}$, we start with the corresponding pairwise error probabilities, which can be derived using the techniques outlined in [3], *i.e.*,

$$P_{\mathcal{PE}_{\mathcal{A}}|\mathbf{H}}, P_{\mathcal{PE}_{\mathcal{B}}|\mathbf{H}} \leq \left[1+\rho\left|h_{1,d}\right|^2+\frac{\rho\left|h_{1,d}\right|^2+\rho\left|b^2 h_{r,d} h_{1,r}\right|^2+\rho^2\left|h_{1,d}\right|^4}{1+\left|b^2 h_{r,d}\right|^2}\right]^{-l/2}, \tag{6}$$

$$P_{\mathcal{PE}_{\mathcal{C}}|\mathbf{H}} \leq \left[1+\rho(\left|h_{1,d}\right|^2+\left|h_{2,d}\right|^2)+\frac{\rho(\left|h_{1,d}\right|^2+\left|h_{2,d}\right|^2)+\rho\left|b^2 h_{r,d}\right|^2(\left|h_{1,r}\right|^2+\left|h_{2,r}\right|^2)}{1+\left|b^2 h_{r,d}\right|^2}\right.$$
$$\left.+\frac{\rho^2(\left|h_{1,d}\right|^2+\left|h_{2,d}\right|^2)^2+\rho^2\left|b h_{r,d}\right|^2\left|h_{1,d} h_{2,r}-h_{2,d} h_{1,r}\right|^2}{1+\left|b^2 h_{r,d}\right|^2}\right]^{-l/2}, \tag{7}$$

where $\mathbf{H}=[h_{1,r}, h_{2,r}, h_{r,d}, h_{1,d}, h_{2,d}]$. Now, $P_{\mathcal{E}_{\mathcal{A}}}$ and $P_{\mathcal{E}_{\mathcal{B}}}$ are obtained by averaging (6) over the ensemble of the channel realizations and further utilizing the union bound over all pairwise error events that lead to $\mathcal{E}_{\mathcal{A}}$ and $\mathcal{E}_{\mathcal{B}}$, *i.e.*,

$$P_{\mathcal{E}_{\mathcal{A}}}, P_{\mathcal{E}_{\mathcal{B}}} \dot{\leq} \rho^{-[(1-r_1)^+ + (1-2r_1)^+]}, \tag{8}$$

where $f(\rho)\dot{\leq}\rho^{-d}$ if $\lim_{\rho\to\infty}\log f(\rho)/\log\rho \leq -d$.

However, averaging to derive $P_{\mathcal{E}_{\mathcal{C}}}$ is not straightforward due to the term

$$\left|h_{1,d} h_{2,r}-h_{2,d} h_{1,r}\right|^2,$$

which involves the subtraction operation. To circumvent this problem, we define

$$\Theta := \frac{h_{2,r} h_{1,d}-h_{1,r} h_{2,d}}{\sqrt{\left|h_{1,r}\right|^2+\left|h_{2,r}\right|^2}} \quad \text{and} \tag{9}$$

$$\Omega := \frac{h_{1,r} h_{1,d}+h_{2,r} h_{2,d}}{\sqrt{\left|h_{1,r}\right|^2+\left|h_{2,r}\right|^2}}. \tag{10}$$

It is then straightforward to see that *conditioned* on $h_{1,r}$ and $h_{2,r}$, $\Theta$ and $\Omega$ are two complex Gaussian random variables with zero mean and unit variance. Furthermore, $E\{\Theta\Omega^*|h_{1,r}, h_{2,r}\}=0$, meaning that $\Theta$ and $\Omega$ are conditionally uncorrelated and therefore independent. Essentially, (9) and (10) can be viewed as a whitening transformation of $h_{1,d}$ and $h_{2,d}$. Realizing that,

$$|\Theta|^2+|\Omega|^2=\left|h_{1,d}\right|^2+\left|h_{2,d}\right|^2,$$

we can rewrite (7) as

$$P_{\mathcal{PE}_{\mathcal{C}}|\mathbf{H}} \leq \left[1+\rho(|\Theta|^2+|\Omega|^2)+\frac{\rho(|\Theta|^2+|\Omega|^2)+\rho\left|b^2 h_{r,d}\right|^2(\left|h_{1,r}\right|^2+\left|h_{2,r}\right|^2)}{1+\left|b^2 h_{r,d}\right|^2}\right.$$
$$\left.+\frac{\rho^2(|\Theta|^2+|\Omega|^2)^2+\rho^2\left|b h_{r,d}\right|^2(\left|h_{1,r}\right|^2+\left|h_{2,r}\right|^2)|\Theta|^2}{1+\left|b^2 h_{r,d}\right|^2}\right]^{-l/2}. \tag{11}$$

Note that in general, $\Theta$, $\Omega$, $h_{1,r}$ and $h_{2,r}$ are correlated. Thus, we cannot directly apply the techniques of [3] to average (11). However, by averaging in two steps, *i.e.*, fixing $h_{1,r}$ and $h_{2,r}$ and taking the average with respect to $\Theta$, $\Omega$ and $h_{r,d}$, and then taking the average with respect to $h_{1,r}$ and $h_{2,r}$, we can characterize $P_{\mathcal{E}_{\mathcal{C}}}$. More specifically, conditioned on $h_{1,r}$ and $h_{2,r}$, we average (11) over the ensemble of codewords and then average with respect to $\Theta$, $\Omega$ and $h_{r,d}$ to obtain,

$$P_{\mathcal{E}_{\mathcal{C}}|h_{1,r},h_{2,r}} \dot{\leq} \rho^{-d_{\mathcal{E}_{\mathcal{C}}|h_{1,r},h_{2,r}}}, \tag{12}$$

where

$$d_{\mathcal{E}_{\mathcal{C}}|h_{1,r},h_{2,r}} = \begin{cases} 2(1-2r_1)^+ & \text{for} \quad \min\{v_{1,r}, v_{2,r}\} > (1-2r_1)^+ \\ [3(1-2r_1) - \min\{v_{1,r}, v_{2,r}\}]^+ & \text{for} \quad 0 \leq \min\{v_{1,r}, v_{2,r}\} \leq (1-2r_1)^+ \end{cases} \tag{13}$$

and $v_{i,r}$ is the corresponding exponential order [2] of $|h_{i,r}|^2$ for $i = 1, 2$.

Averaging (14) with respect to $v_{1,r}$ and $v_{2,r}$, it then follows that

$$P_{\mathcal{E}_{\mathcal{C}}} \dot{\leq} \rho^{-3(1-2r_1)^+}. \tag{14}$$

Now, (14) together with (8) and (5) results in (4), and thus completes the proof. ■

## III. Discussion

For purposes of comparison, we first recall an upper bound on the achievable DMT in the symmetric MARC, along with the DMT's of the DDF and CF protocols. For the symmetric MARC with two users and one relay, an upper bound on the achievable DMT for $0 \leq r_1 \leq 1/2$ is [5],

$$d_{MARC}(r_1) \leq \begin{cases} 2 - 2r_1, & \text{for } 0 \leq r_1 \leq \frac{1}{4} \\ 3(1-2r_1), & \text{for } \frac{1}{4} \leq r_1 \leq \frac{1}{2} \end{cases}. \tag{15}$$

On the other hand, the DMT of DDF for $0 \leq r_1 \leq 1/2$ is [5],

$$d_{DDF}(r_1) = \begin{cases} 2 - 2r_1, & \text{for } 0 \leq r_1 \leq \frac{1}{4} \\ 3(1-2r_1), & \text{for } \frac{1}{4} \leq r_1 \leq \frac{1}{3} \\ \frac{1-2r_1}{r_1}, & \text{for } \frac{1}{3} \leq r_1 \leq \frac{1}{2} \end{cases}, \tag{16}$$

[2]Assume $h$ is a random variable, its exponential order $v := \frac{\log|h|^2}{\log \rho}$

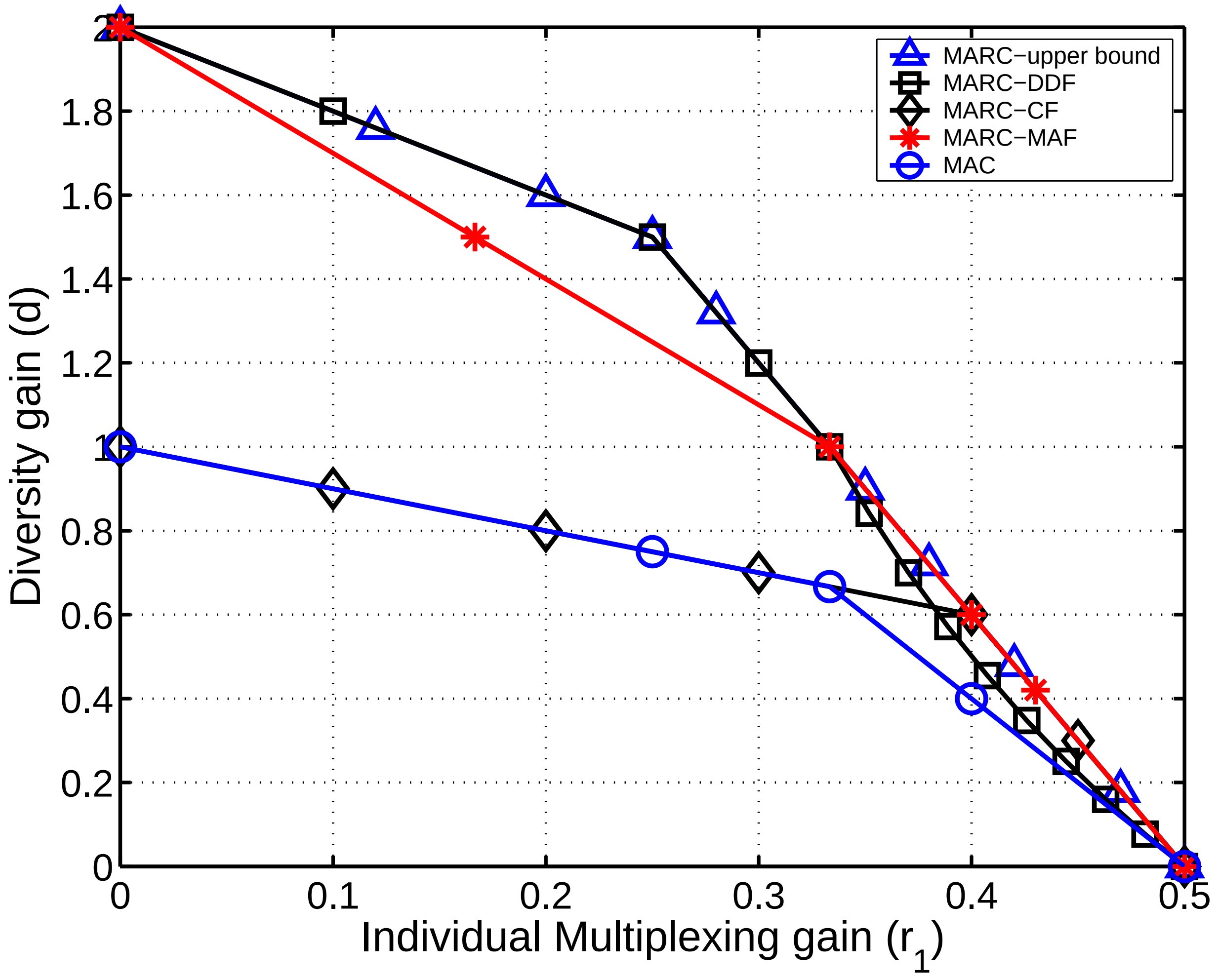

Fig. 2. DMT of the different MARC protocols.

and that of CF for $0 \leq r_1 \leq 1/2$ is [6],

$$d_{CF}(r_1) = \begin{cases} 1 - r_1, & \text{for } 0 \leq r_1 \leq \frac{2}{5} \\ 3(1 - 2r_1), & \text{for } \frac{2}{5} \leq r_1 \leq \frac{1}{2} \end{cases}. \tag{17}$$

To highlight the advantage gained from adding a single relay, we also recall the DMT of a symmetric MAC with two users [10],

$$d_{MAC}(r_1) = \begin{cases} 1 - r_1, & \text{for } 0 \leq r_1 \leq \frac{1}{3} \\ 2(1 - 2r_1), & \text{for } \frac{1}{3} \leq r_1 \leq \frac{1}{2} \end{cases}. \tag{18}$$

These trade-offs, along with the trade-off for MAF in Theorem 1, are shown in Fig. 2. From the results and figure, we make the following observations:

1) The MAF protocol achieves the optimal DMT for $1/3 \leq r_1 \leq 1/2$. In fact, over this range of multiplexing gains, MAF behaves like a MISO system with three transmit antennas and one receive antenna.
2) MAF uniformly dominates the CF protocol in terms of DMT, *i.e.*, $\forall r_1, d_{MAF}(r_1) \geq d_{CF}(r_1)$. Relative to MAF, CF suffers from a significant loss in diversity gain at low multiplexing gain. In particular, MAF achieves the full diversity gain $2$ as $r_1$ vanishes to $0$; in contrast, CF only achieves a diversity gain $1$ as $r_1$ vanishes to $0$. Compared to CF, MAF enjoys another advantage of lower complexity at the relay.
3) It is somewhat surprising to observe that MAF outperforms DDF in terms of DMT for $1/3 \leq r_1 \leq 1/2$, considering that AF relay protocols generally suffer from a significant performance loss in the high multiplexing regime for the half-duplex relay channel [2], [3]. An intuitive explanation for this observation will be provided in the sequel.
4) In the regime of $1/3 \leq r_1 \leq 2/5$, neither DDF nor CF is optimal; but MAF is. To the best of our knowledge, MAF is the only protocol that achieves the optimal DMT in this regime.
5) Even over the range of multiplexing gains for which MAF becomes suboptimal, *i.e.*, $0 \leq r_1 \leq 1/3$, the achieved DMT is identical to that of the NAF relay [3] with a single user. In other words, for low multiplexing gains, each user benefits from the relay as if it was the only user present. Also, in this regime, the DMT gap between DDF and MAF is much smaller compared to the gap between DDF and CF.
6) The DMT of MAF uniformly dominates that of MAC and reveals the tremendous advantage that a number of users could potentially gain from a single MAF relay. The DMT of DDF approaches that of MAC in the high multiplexing regime. Thus, the gain of a complicated DDF relay diminishes in the regime of high multiplexing. The DMT of CF overlaps with that of MAC in the regime of low multiplexing gains. This implies that there may be no advantage of employing a CF relay for a number of users when the multiplexing gain is small.

The surprising advantage of MAF over DDF at high multiplexing gains can be attributed to the following two factors. First, at high multiplexing gains, DDF might require the relay to spend a large percentage of time decoding the two users' messages. As a result, the relay may not have enough time to retransmit them. Second, compared to the NAF relay protocol, the correlation between the two halves of the overall signal received at the destination in MAF is reduced. As a result, the overall signal resembles a repetition code to a lesser extent and consequently, the performance is improved. In fact, based on this rationale, one might expect MAF to become optimal for progressively larger ranges of multiplexing gains, as the number of users increases.

We conclude this section by presenting simulation results for MAF and DDF at different multiplexing gains. Fig. 3 shows the outage probabilities $P_{\mathcal{O}}(R)$ of DDF and MAF with $R = r_1 \log(1+\rho)$. When $r_1 = 0.2$, the outage probability curve of DDF demonstrates a steeper slope compared to that of MAF, indicating a higher diversity gain for DDF. However, for $r_1 = 0.4$, the intersection between the curve of MAF and that of DDF suggests that MAF has a higher diversity gain. These observations from simulations are in line with what we predict from the DMT, *i.e.*, the diversity gain of MAF is higher than that of DDF at high multiplexing gains, but is smaller at low multiplexing gains. They also suggest that complete system design requires characterization of not only the DMT, which captures the exponential behavior of the error probability with SNR, but also the leading coefficients that capture the geometric dependence and ”coding gain” of the relaying protocols.

## IV. Conclusion

Because several previous works on the muti-access relay channel (MARC) have focused on protocols that requires complicated signal processing at the relay [5], [6], this paper's main contribution is to proposes a linear relaying protocol, *i.e.*, multi-access amplify-forward (MAF), which not only reduces complexity of relaying but also achieves good performance in slow fading environments. MAF achieves the optimal diversity-multiplexing trade-off in the high multiplexing regime. In particular, in the regime of $1/3 \leq r_1 \leq 2/5$, for which neither DDF nor CF is optimal, MAF achieves the optimal diversity-multiplexing trade-off. In the low multiplexing regime, MAF allows each user to gain cooperative diversity as if there is no interference from other users and no contention for the relay. Compared with other protocols, *e.g.*, DDF and CF, MAF achieves good performance at a low complexity and can be an appealing architectural alternative to architectures that exploit user cooperation.

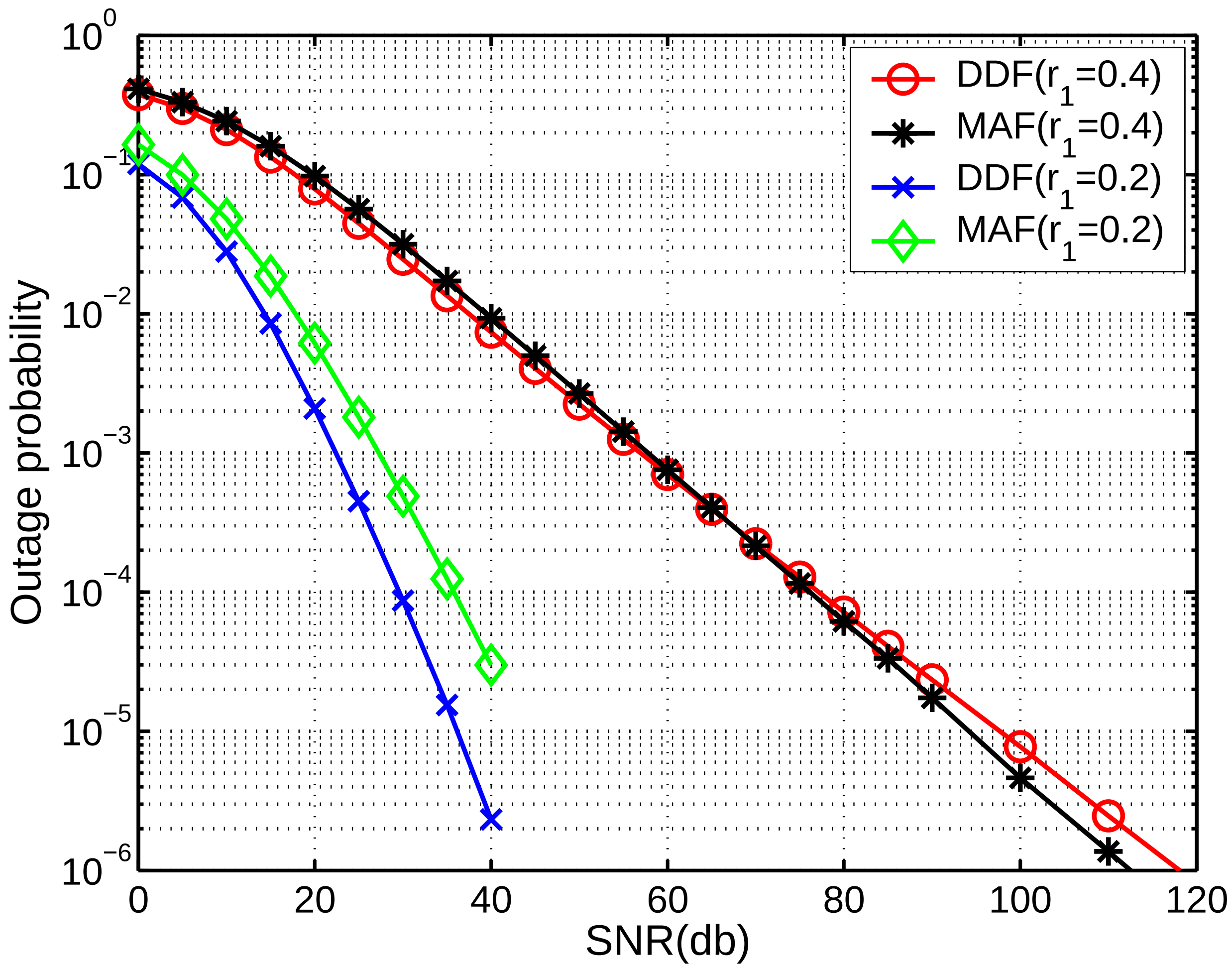

Fig. 3. Outage probabilities $P_{\mathcal{O}}(R)$ for DDF and MAF. Note that $R = r_1 \log(1+\rho)$